\documentstyle[12pt]{article}

\input{tcilatex}

\begin{document}

\title{Creating Boolean Functions for the Five-EPR-Pair,
Single-Error-Correcting Code}
\author{Jin-Yuan Hsieh$^{1}$ and Che-Ming Li$^{2}$ \\
$^{1}$Department of Mechanical Engineering, Ming Hsin University \\
of Science and Technology, Hsinchu 30401,Taiwan.\\
$^{2}$Institute and Department of Electrophysics, National Chiao\\
Tung University, Hsinchu 30050, Taiwan.}
\maketitle

\begin{abstract}
A quantum single-error-correcting scheme can be derived from a one-way
entanglement purification protocol in purifying one Bell state from a finite
block of five Bell states. The main issue to be concerned with in the theory
of such an error-correction is to create specific linear Boolean functions
that can transform the sixteen error syndromes occurring in the
error-correcting code onto their mappings so that one Bell state is
corrected whenever the other four in the finite block are measured. The
Boolean function is performed under the effect of its associated sequence of
basic quantum unilateral and bilateral operations. Previously, the Boolean
function is created in use of the Monte Carlo computer search method. We
introduce here a systematic scenario for creating the Boolean function and
its associated sequence of operations so that we can do the job in an
analytical way without any trial and error effort. Consequently, all
possible Boolean functions can in principle be created by using our method.
Furthermore, for a deduced Boolean function, we can also in the spirit \ of
our method derive its best associated sequence of operations which may
contain the least number of total operations or the least number of the
bilateral XOR operations alone. Some results obtained in this work show the
capability of our method in creating the Boolean function and its sequence
of operations.

\end{abstract}

\section{Introduction}

Entanglement plays an important role in quantum information processing for
transmitting unknown quantum states via noisy channels from a sender to a
receiver, such as quantum teleportation\cite{1}, quantum data compression%
\cite{2}, and quantum super-dense coding\cite{3}, etc.. To achieve a
reliable transmission of the unknown states, pure maximally entangled pairs,
typically the Einstein-Podolsky-Rosen (EPR) pairs, or the Bell-states, which
are emerged from a quantum resource and transmitted through a quantum
channel, should be shared by the sender (Alice) and the receiver (Bob).
Because of the presence of noise in the quantum channel, Alice and Bob
therefore have to perform actions such as entanglement purification
procedures in distilling pure entangled states from a larger number of
impure entangled states. The entanglement purification protocol (EPP) allows
Alice and Bob to perform local unitary transformations and measurements and
even allows them to coordinate their actions through one-way or two-way
classical communication. It, however, does not allow Alice and Bob to
perform non-local actions nor to transmit fresh quantum states from one to
the other. An EPP involving two-way communication is called a two-way EPP
(2-EPP), in which both Alice and Bob need to know the results of measurement
from each other. Typical 2-EPPs include the IBM protocol\cite{4} and the
Oxford protocol\cite{5}, which also belong to the recurrence method. On the
other hand, a one-way EPP (1-EPP) requires only Alice to send her
measurement result through classical channel to Bob, who when combining it
with his own result can decide a following action to perform. Thus, the
1-EPP can produce pure maximally entangled pairs which are separated both in
space and in time. The hashing protocol\cite{6} and the breeding protocol%
\cite{4} are examples of the 1-EPP.

Normally, the 2-EPP can be combined with the 1-EPP, such as a
recurrence-hashing protocol, to produce a higher purification yield, defined
by a ratio $m/n$, where $m$ is the number of the purified useful pairs and $%
n $ is the number of the input impure pairs. It is well known that if the
final fidelity of the purified states $F\rightarrow 1$, i.e., if the the
final state is almost of the wanted pure state, is required, then the
initial number of the impure pairs should be $n\rightarrow \infty $ for both
the 1-EPP and 2-EPP. As shown by Bennett et al.\cite{6}, the pure one way
hashing and breeding protocols can produce a non-zero yield only when the
fidelity of the purified states possessed by the input impure pairs is
greater than $F\approx 0.8107$. Even the best 1-EPP, proposed by Shor and
Smolin\cite{7}, can do the job only when the initial fidelity of the
purified states exceeds $F\approx 0.8049$. When the 2-EPPs are performed, on
the other hand, the input state becomes distillable if its fidelity of the
purified states is greater than $F=0.5$\cite{4}\cite{5}. So , in these
senses, the 2-EPPs perform better than the 1-EPP when they are only used in
conjunction with teleportation in offering a means of transmitting quantum
information through noisy channels. The 1-EPP, however, by producing
time-separated entanglement, can additionally be used to protect quantum
states during storage in a noisy environment. This important feature then
leads to the consequence that the 1-EPP, when combined with quantum
teleportation, can always permit the creation of a quantum error-correcting
code (QECC). Bennett et al.\cite{6} have presented the equivalence between
the 1-EPP and QECC. The QECC derived from a 1-EPP will have a code-rate
equal to the yield of the 1-EPP and have a fidelity equal to the fidelity of
the purified states produced by the 1-EPP. The one-way hashing protocol
therefore can be interpreted as an error-correcting code which protects an
arbitrary state in $2^{m}$-dimensional Hilbert space from noise by employing
qubit block of asymptotically large size $n$. It was shown\cite{6} that the
QECC derived from the one-way hashing protocol actually does its job better
than the QECCs based on the linear-code theory of Calderbank and Shor\cite{8}
and Steane\cite{9} in the sense that it has a higher rate than the latter
ones. Nevertheless, as most of the QECCs [8-16] focused on employing qubits
of finite block size $n$ in protecting $m<n$ qubits from any error on no
more than $t$ qubits, Bennett et al.\cite{6} have further discussed the
possibility of transforming the 1-EPP into a QECC in the same manner. They
ended up with a QECC by showing how finite blocks of 5 EPR pairs can be
purified in the presence of noise which only affects at most one of the Bell
states. The theory of quantum error-corrections independently presented by
Knill and Laflamme\cite{17} has proved the block size $n=5$ is the smallest
and best case. So far, a 5-qubit, single-error-correcting code has been
performed experimentally\cite{18}.

In the 5-EPR pair, single-error-correcting quantum block code, there are
totally 16 possible blocks of Bell state (i.e., 16 error syndromes) to be
dealt with; one corresponds to the Bell states without error and the
remaining ones are those with one of three errors on only one of the five
Bell states. When the block of Bell states is coded as a string of
phase-amplitude bits, the quantum error-correction problem then turns out to
create a classical Boolean function which can map exactly one on one the
syndrome states onto others such that the first Bell state (if the five Bell
states are numbered in order) is always the same when the remaining four are
measured and found out to be the same. Associated with the Boolean function,
there is a sequence of basic operations including uni-lateral and bi-lateral
rotations acting on one Bell state and bi-lateral XORs (controlled-NOTs)
acting on two Bell states to realize the QECC protocol. In fact, there are
so many possible Boolean functions, and therefore so many corresponding
sequences of operations, for such QECC protocols; some of them are more
economic and feasible than the others because they need the least operations
to complete the QECC procedures. Bennett et al.\cite{6} have introduced the
Monte Carlo method to numerically find out possible Boolean functions by
randomly choosing the sequence of operations acting on randomly chosen
states. Implementing the numerical Mote Carlo method, however, is not a
systematic way and therefore is hard to find out all possible Boolean
functions to complete the theory of the 5-EPR-pair QECC. Based on this
reason, we are intended in this work to present a systemic method for
establishing the Boolean functions and their corresponding sequences of
operations for such QECCs.

The Mote Carlo method is a forward way of serial trial and error, i.e.,
during the procedure of creating the Boolean function, the mapped states of
the $16$ syndromes are consecutively checked to see if the Boolean words for
measurement embedded in them are independent. In our method, an array of the
independent Boolean words for measurements is prescribed, and then in the
opposite direction of the Monte Carlo method, Boolean functions are
established consistently in an easy way very similar to the transformation
of an invertible matrix into the identity matrix by performing elementary
row operations on the matrices following the theory of linear algebra. Since
the transformation of a matrix is physically controlled by the operations
mentioned above, the sequence of operations associated with the established
Boolean function can be decided accordingly. It is therefore very easy to
find out a Boolean function and its corresponding sequence of operations
analytically and all possible solutions according to the prescribed array of
Boolean words for measurement can be found if aided by a numerical program.
In the next section, we will describe the derivation of the QECC from the
1-EPP used to purify a Bell state from a finite block of 5 Bell states. In
section 3, we then describe in detail the systematic method for creating
Boolean functions in the QECC. A typical example of the Boolean function is
given to help interpreting the procedure of our method. In section 4, more
results are given to show the capability of the present method and a brief
discussion is also presented. A conclusion is given in the final section.

\section{The 5-EPR-pair single-error-correcting code}

The single-error-correcting code considered herein is schematically shown in
Fig. 1, which plots the combination of the 1-EPP that purify finite blocks
of five EPR pairs and the teleportation that safely transmit arbitrary
states. Alice first prepares mixed states $\widehat{{\bf M}}$ by passing
halves of one standard state among the Bell states $\Phi ^{\pm }=(\left|
00\right\rangle \pm \left| 11\right\rangle )/\sqrt{2}$ and $\Psi ^{\pm
}=(\left| 01\right\rangle \pm \left| 10\right\rangle )/\sqrt{2}$ from source
E and through noisy channel. For convenience, the state $\Phi ^{+}$ is
considered as the standard Bell state in this work. The mixed state $%
\widehat{{\bf M}}$ is Bell diagonal under the restriction that the noise
model is one-sided (i.e., N$_{\text{A}}$ is absent), or effectively one-sided%
\cite{6}. (In fact, any noise can be made effectively one-sided as a
twirling operation performed by Alice and Bob can convert any bipartite
mixed state into a Bell diagonal states or a Werner state.) Alice and Bob
then perform the 1-EPP to yield perfectly entangled states (*, which may be
singlets $\Psi ^{-}$ or any one of the triplets $\Psi ^{+}$ and $\Phi ^{\pm
} $) used to teleport an arbitrary state $\left| \xi \right\rangle $ safely
from Alice to Bob, completing a QECC. In developing the theory of the
5-EPR-pair single-error-correcting code first introduced by Bennett et al.%
\cite{6}, the four Bell states are first coded as classical two-bit words,
which read

\begin{equation}
\Phi ^{+}=00,\text{ }\Phi ^{-}=10,\text{ }\Psi ^{+}=01,\text{ }\Psi ^{-}=11.
\end{equation}%
Here the left bits are high-order or phase bits used to identify the $+/-$
property and the right bits are low-order or amplitude bits used to identify
the $\Phi /\Psi $ property. Consequently, each finite block of five Bell
states used in the present QECC then can be represented by a ten-bit word,
for example, $\Phi ^{+}\Psi ^{-}\Phi ^{+}\Phi ^{+}\Phi ^{+}=0011000000.$
Since only single error is allowed in the present model, and the error could
be either a phase error ($\Phi ^{+}\rightarrow \Phi ^{-}$), an amplitude
error ($\Phi ^{+}\rightarrow \Psi ^{+}$), or both ($\Phi ^{+}\rightarrow
\Psi ^{-}$)\cite{11}\cite{13}, therefore, after the action of noise N$_{%
\text{B}}$, there is a set of totally 16 possible ten-bit words defining the
complete set of error syndromes to be dealt with. The error syndrome words
are denoted by 16 Boolean valued ( $\in \{0,1\}$) vectors $x^{(i)}$, $%
i=0,1,2,...,15,$ in a ten-dimensional vector space. Clearly, if we denote
the null vector in the vector space by $x^{(0)}=00...00$, which represents
the no-error state, and the three states in which the single error occurs on
Bell state $k$, $k=1,2,...,5$, by $x^{(3k-2)}$, $x^{(3k-1)}$, and $x^{(3k)}$%
, respectively, then the 16 error syndrome vectors can be subdivided into 5
four-groups denoted by $V_{x}^{(k)}=\{x^{(0)},x^{(3k-2)},x^{(3k-1)},x^{(3k)}%
\}$, where $x^{(0)}$ is the identity element and the property $%
x^{(3k-2)}\oplus x^{(3k-1)}=x^{(3k)}$ holds ($\oplus $ is the addition
modulo 2). Excluding the null vector $x^{(0)}$, two distinct elements
arbitrarily chosen from each of the 5 four-groups $V_{x}^{(k)}$ are
independent to those chosen from the other four-groups, so 10 independent
vectors can be chosen from the 5 four-groups to form a set of basis vectors
spanning the 10-dimensional Boolean valued vector space. Meanwhile, all
four-bit Boolean valued vectors, totally 16 and each denoted by $v^{(i)}$ in
accord with error syndrome $i$, can also be subdivided into 5 four-groups $%
V_{v}^{(k)}=\{v^{(0)},v^{(3k-2)},v^{(3k-1)},v^{(3k)}\}$, which are
one-to-one isomorphic to the four-groups $V_{x}^{(k)}$ in the same manner
that $v^{(0)}=0000$ and

\begin{equation}
v^{(3k-2)}\oplus v^{(3k-1)}=v^{(3k)},\text{ }k=1,2,...,5.
\end{equation}%
One example of the correspondence between the four-groups $V_{v}^{(k)}$ and $%
V_{x}^{(k)}$ is shown in Table 1.

The 5-EPR-pair, single-error-correcting code then demands Alice and Bob to
apply unitary transformations $U_{1}$ and $U_{2}$ in mapping $x^{(i)}$ onto
other two-bit words $w^{(i)}$ which, when four of the five Bell states in a
block are locally measured, should yield results represented by $v^{(i)}.$
Thus the 16 vectors $w^{(i)}$ should also be subdivided into 5 four-groups $%
V_{w}^{(k)}$ accordingly. The measurement results should be distinguished by
reading the low bits of the measured Bell states so the ten-bit word $%
w^{(i)} $ in fact has its four low bits of the measured Bell states denoted
by $v^{(i)}$. Without loss of generality, in what follows the four
components of $v^{(i)}$ are located at the 4$^{th},$ 6$^{th},$ 8$^{th},$ and
10$^{th}$ components of $w^{(i)}$ , respectively. The state of the remaining
unmeasured EPR pair then is represented by a truncated word $w^{^{\prime
}(i)}$, the first two bits of the vector $w^{(i)},$ for every error syndrome 
$i$. A successful error-correction should require that, whatever $%
w^{^{\prime }(i)}$ is, Bob can always perform a corresponding rotation $%
U_{3}^{(i)}$ on it and restore it to the standard state $\Phi ^{+}=00$, or
any one of the other Bell states, after learning the measurement results $%
v^{(i)}.$ The unilateral rotation $U_{3}^{(i)}$ performed by Bob can be
either one element of the Pauli group $\{{\bf 1},$ $\sigma _{x},$ $\sigma
_{y},$ $\sigma _{z}\},$ which , if the remaining unmeasured Bell state is to
be restored to the standard $\Phi ^{+}$, for example, can do the
transformations

\begin{eqnarray}
{\bf 1} &{\bf :}&\text{ }\Phi ^{+}(00)\rightarrow \Phi ^{+}(00),  \nonumber
\\
\sigma _{x} &:&\text{ }\Psi ^{+}(01)\rightarrow \Phi ^{+}(00),  \nonumber \\
\sigma _{y} &:&\text{ }\Psi ^{-}(11)\rightarrow \Phi ^{+}(00),  \nonumber \\
\sigma _{z} &:&\text{ }\Phi ^{-}(10)\rightarrow \Phi ^{+}(00).
\end{eqnarray}%
respectively.

The unmeasured Bell states, coded by $w^{\prime (i)}$, and the corresponding
Pauli rotations $U_{3}^{(i)}$, can be decided by Bob after he has learned
the measurement results $v^{(i)},$ if both Alice and Bob have pre-agreed
with prescribed unitary transformation $U_{1}$ and $U_{2}$, which represent
a sequence of unilateral and bilateral operations performing the
transformations $x^{(i)}\rightarrow w^{(i)}$. In EPPs, and in the QECCs
derived from them, a Bell state is transformed into another under some
typical unitary operations performed by either Alice or Bob, but not both,
and bilateral operations that require both Alice and Bob to perform a same
transformation on their spins. The typical unilateral operations include the
Pauli transformations $\sigma _{x},$ $\sigma _{y},$ and $\sigma _{z}$, which
perform a $\pi $ rotation of Alice or Bob's spin about the $x$-, $y$-,and $z$%
-axis, respectively. The typical bilateral operations, on the other hand,
can be either the operations denoted by $B_{x}$, $B_{y}$, and $B_{z}$ that
require both Alice and Bob to perform a $\pi /2$ rotation of their spins in
an EPR pair about the $x$-, $y$-,and $z$-axis, respectively, or a bilateral
XOR (BXOR) that requires both Alice and Bob to perform a controlled NOT
operation on their spins in common source and target pairs. In the present
5-EPR-pair QECC, however, Alice and Bob are only confined to performing a
sequence of four particular operations which can do anything required in the
transformations $x^{(i)}\rightarrow w^{(i)}$. The four basic operations
include: (1) a BXOR, which, in our classical bit representation, performs
the transformation $(x_{S}$, $y_{S})(x_{T},y_{T})\rightarrow (x_{S}\oplus
x_{T}$, $y_{S})(x_{T},y_{S}\oplus y_{T})$, where the subscripts $S$ and $T$
denote the source and target pairs, respectively; (2) a bilateral $\pi /2$
rotation $B_{y}$, which performs $(x,$ $y)\rightarrow (y,x)$; (3) a
composite operation $\sigma _{x}B_{x}$, which performs $(x,$ $y)\rightarrow
(x,x\oplus y)$; and (4) a unilateral $\pi $ rotation $\sigma _{z}$, which
complements the high bit of an EPR pair, viz., $(x,$ $y)\rightarrow (x\oplus
1,y).$

The effect of such a sequence of the four basic operations mentioned above
is to apply a linear Boolean function mapping $x^{(i)}$ onto $w^{(i)}$ ,
which can \ also be written by a matrix equation

\begin{equation}
w^{(i)}={\bf M}_{wx}x^{(i)}\oplus b\text{,}
\end{equation}%
where ${\bf M}_{wx}$ is a $10\times 10$ inversible matrix (i.e., $\det
(M_{wx})=1$) defined in the 10-dimensional Boolean valued vector space and
the Boolean valued vector $b$ corresponds to the mapping of the null vector $%
x^{(0)}$, $b=w^{(0)}$. Without loss of generality, in what follows, $%
b=00...00$ is assumed so that the unilateral operation $\sigma _{z}$ can be
excluded from the sequence of operations, i.e., the Boolean function is
reduced to be

\begin{equation}
w^{(i)}={\bf M}_{wx}x^{(i)}\text{,}
\end{equation}%
where the null vector $x^{(0)}$ thus remains unchanged under the
transformation. Usually, an $n-$dimensional vector is represented by an $%
n\times 1$ column matrix, then eq. (5) in fact implies

\begin{equation}
{\bf M}_{w}={\bf M}_{wx}{\bf I}_{x},
\end{equation}%
with

\begin{equation}
{\bf M}_{w}=\left[
w^{(1)}w^{(2)}w^{(4)}w^{(5)}w^{(7)}w^{(8)}w^{(10)}w^{(11)}w^{(13)}w^{(14)}%
\right] ,
\end{equation}

\begin{equation}
{\bf I}_{x}=\left[
x^{(1)}x^{(2)}x^{(4)}x^{(5)}x^{(7)}x^{(8)}x^{(10)}x^{(11)}x^{(13)}x^{(14)}%
\right] ,
\end{equation}%
where $w^{(3k-2)}$, $w^{(3k-1)}\in V_{w}^{(k)}$ and $x^{(3k-2)}$, $%
x^{(3k-1)}\in V_{x}^{(k)}$, $k=1,2,...,5$. Note that we can always convert
the matrix ${\bf I}_{x}$ (8) to the $10\times 10$ identity matrix ${\bf I}%
_{10}$, by adding columns to columns (in their associated four-groups) and
interchanging columns in $I_{x}$. Since each column of the matrix ${\bf M}%
_{w}$ (7) should have its $4^{th}$, $6^{th}$, $8^{th}$, and $10^{th}$
components forming a measurement-result vector, an array for the measurement
results can be written by

\begin{eqnarray}
{\bf M}_{v} &=&\left[
v^{(1)}v^{(2)}v^{(4)}v^{(5)}v^{(7)}v^{(8)}v^{(10)}v^{(11)}v^{(13)}v^{(14)}%
\right]  \nonumber \\
&=&{\bf M}_{vw}{\bf M}_{w},
\end{eqnarray}%
where

\begin{equation}
{\bf M}_{vw}=\left[ 
\begin{array}{cccccccccc}
0 & 0 & 0 & 1 & 0 & 0 & 0 & 0 & 0 & 0 \\ 
0 & 0 & 0 & 0 & 0 & 1 & 0 & 0 & 0 & 0 \\ 
0 & 0 & 0 & 0 & 0 & 0 & 0 & 1 & 0 & 0 \\ 
0 & 0 & 0 & 0 & 0 & 0 & 0 & 0 & 0 & 1%
\end{array}%
\right] .
\end{equation}%
In order to achieve a successful error-correction, the ten
measurement-result vectors $v^{(3k-2)}$ and $v^{(3k-1)}$ appearing in ${\bf M%
}_{v}$ (9), the distinct five vectors $v^{(3k)}$ derived from eq. (2), and
the null vector $v^{(0)}$ should form the complete 4-dimensional Boolean
valued vector space. Therefore, a suitable array of the measurement-result
vectors can be given by

\begin{equation}
{\bf M}_{v}=\left[ 
\begin{array}{cccccccccc}
1 & 0 & 0 & 1 & 0 & 0 & 0 & 1 & 1 & 0 \\ 
0 & 0 & 1 & 0 & 1 & 1 & 1 & 1 & 1 & 0 \\ 
0 & 1 & 0 & 0 & 0 & 1 & 1 & 0 & 1 & 0 \\ 
0 & 0 & 0 & 1 & 1 & 0 & 1 & 0 & 0 & 1%
\end{array}%
\right] ,
\end{equation}%
for instance.

The main issue in the theory of the present QECC now turns out to be the
creation of a Boolean function $x^{(i)}\rightarrow w^{(i)}$, or simply the
creation of the matrix ${\bf M}_{w}$ (7), the effect of a specific sequence
of the basic operations BXOR, $B_{y}$, and $\sigma _{x}B_{x}$. To create
Boolean functions, Bennett et al.\cite{6} performed a Monte Carlo computer
search for the corresponding sequence of the four basic operations
representing the actions of unitary transformations $U_{1}$ and $U_{2}$.
Their program randomly selects one of the four basic operations and randomly
selects a Bell state or pair of Bell states to which to apply the operation.
Then the program checks if the resulting set of states $w^{(i)}$ results in
the success of an error-correction; if not, the program then repeats the
procedure by adding another random operation. Basically, the approach that
Bennett et al. implemented is a tedious numerical method of trial and error
performing the transformation ${\bf I}_{x}\rightarrow {\bf M}_{w}$ subject
to a ''forward'' sequence of the four basic operations. In this work, we
will present an analytical method for creating Boolean functions implemented
in the present QECC. The method to be introduced is in fact an inverse way
to the Monte Carlo method. In the method, the inverse transformation ${\bf M}%
_{w}\rightarrow {\bf I}_{x}$\ is considered to be performed subject to a
''backward'' sequence of operations, which is exactly in reverse order of
the ''forward'' one corresponding to the transformation ${\bf I}%
_{x}\rightarrow {\bf M}_{w}$ \ because each basic operation used in the QECC
is its own inverse operation. The present method will be described in detail
in the following section.

\section{The present method}

In deducing the analytical method for creating Boolean functions for the
5-EPR-pair, single -error-correcting code, the $10\times 10$ matrix ${\bf M}%
_{w}$ is sometimes rewritten in an alternative form of a $5\times 5$ matrix
whose elements are $2\times 2$ matrices, namely, the matrix can be expressed
by

\begin{equation}
{\bf M}_{w}=\left[ 
\begin{array}{cccc}
m_{11} & m_{12} & \cdots & m_{15} \\ 
m_{21} & m_{22} & \cdots & m_{25} \\ 
\vdots & \vdots &  & \vdots \\ 
m_{51} & m_{52} & \cdots & m_{55}%
\end{array}%
\right] ,
\end{equation}%
where the rows enumerate the five Bell states in a block and the columns
correspond to the five four-groups $V_{w}^{(k)}$, respectively. Similarly,
the original $10\times 10$ matrix ${\bf I}_{x}$ can also be rewritten in a
form of $5\times 5$ matrix, such that each column and each row of the $%
5\times 5$ matrix will have one $2\times 2$ element of determinant 1 and
four $2\times 2$ zero matrices. Now, if an array of the measurement result
vectors suitable for a successful error-correction is prescribed, the job in
principle is to perform the transformation ${\bf M}_{w}\rightarrow {\bf I}%
_{x}$ using elementary row operations on the matrix subject to the effect of
a sequence of the basic operations BXOR, $B_{y}$, and $\sigma _{x}B_{x}$.

The first step of the present method is to designate a suitable array of the
measurement result vectors $v^{(i)}$, in which the five four-groups $%
V_{v}^{(k)}$ are constructed. In this work, the array of $v^{(i)}$ shown in
(11) is taken as the designation and then a suitable matrix ${\bf M}_{w}$ is
assumed for a successful error-correction. The assumed matrix is written by

\begin{equation}
{\bf M}_{w}=\left[ 
\begin{array}{cccccccccc}
a_{1} & a_{2} & a_{3} & a_{4} & a_{5} & a_{6} & a_{7} & a_{8} & a_{9} & 
a_{10} \\ 
b_{1} & b_{2} & b_{3} & b_{4} & b_{5} & b_{6} & b_{7} & b_{8} & b_{9} & 
b_{10} \\ 
c_{1} & c_{2} & c_{3} & c_{4} & c_{5} & c_{6} & c_{7} & c_{8} & c_{9} & 
c_{10} \\ 
1 & 0 & 0 & 1 & 0 & 0 & 0 & 1 & 1 & 0 \\ 
d_{1} & d_{2} & d_{3} & d_{4} & d_{5} & d_{6} & d_{7} & d_{8} & d_{9} & 
d_{10} \\ 
0 & 0 & 1 & 0 & 1 & 1 & 1 & 1 & 1 & 0 \\ 
e_{1} & e_{2} & e_{3} & e_{4} & e_{5} & e_{6} & e_{7} & e_{8} & e_{9} & 
e_{10} \\ 
0 & 1 & 0 & 0 & 0 & 1 & 1 & 0 & 1 & 0 \\ 
f_{1} & f_{2} & f_{3} & f_{4} & f_{5} & f_{6} & f_{7} & f_{8} & f_{9} & 
f_{10} \\ 
0 & 0 & 0 & 1 & 1 & 0 & 1 & 0 & 0 & 1%
\end{array}%
\right] ,
\end{equation}
where, in the representation of $5\times 5$ matrix, the $2\times 2$ elements
are

\[
m_{11}=\left[ 
\begin{array}{cc}
a_{1} & a_{2} \\ 
b_{1} & b_{2}%
\end{array}%
\right] ,\text{ }m_{21}=\left[ 
\begin{array}{cc}
c_{1} & c_{2} \\ 
1 & 0%
\end{array}%
\right] ,\text{ ...,} 
\]%
and so forth. Here all the unknowns $a_{r}$, $b_{r}$, ..., $f_{r}$, $r=1,$ $%
2,$ ... $10$, are Boolean valued. The next step of our method is a procedure
of elementary row operations on the matrix ${\bf M}_{w}$ (13), subject to a
suitable sequence of the basic operations. When the assumed matrix ${\bf M}%
_{w}$ is transformed into a matrix ${\bf I}_{x}$\ under the series of row
operations, the unknowns $a_{r}$, $b_{r}$, ..., $f_{r}$ will be solved
stepwise in accord with the structure of ${\bf I}_{x}.$\ It is easy to show
a sequence of row operations can do the transformation on, say, Bell states $%
\alpha $ and $\beta $ in a four-group enumerated by $\gamma $,

\begin{equation}
\left[ 
\begin{array}{c}
m_{\alpha \gamma } \\ 
m_{\beta \gamma }%
\end{array}%
\right] \rightarrow \left[ 
\begin{array}{c}
e \\ 
0%
\end{array}%
\right] ,
\end{equation}%
provided that $\det (m_{\alpha \gamma })=1$ and $\det (m_{\beta \gamma })=0.$
Here $e$ denotes any one of the six possibilities for a $2\times 2$ matrix
whose determinant is unity. ( $m_{\alpha \gamma }$\ in fact should belong to
one of the possibilities.) For example, the consecutive transformation

\[
\left[ 
\begin{array}{c}
m_{\alpha \gamma } \\ 
m_{\beta \gamma }%
\end{array}%
\right] =\left[ 
\begin{array}{c}
\begin{array}{cc}
1 & 0 \\ 
0 & 1%
\end{array}
\\ 
\begin{array}{cc}
1 & 1 \\ 
0 & 0%
\end{array}%
\end{array}%
\right] \rightarrow \left[ 
\begin{array}{c}
\begin{array}{cc}
1 & 0 \\ 
0 & 1%
\end{array}
\\ 
\begin{array}{cc}
0 & 0 \\ 
1 & 1%
\end{array}%
\end{array}%
\right] \rightarrow \left[ 
\begin{array}{c}
\begin{array}{cc}
1 & 0 \\ 
1 & 1%
\end{array}
\\ 
\begin{array}{cc}
0 & 0 \\ 
1 & 1%
\end{array}%
\end{array}%
\right] \rightarrow \left[ 
\begin{array}{c}
\begin{array}{cc}
1 & 0 \\ 
1 & 1%
\end{array}
\\ 
\begin{array}{cc}
0 & 0 \\ 
0 & 0%
\end{array}%
\end{array}%
\right] 
\]%
can be accomplished if the operation $B_{y}$ is first performed on Bell
state $\beta $, then a $\sigma _{x}B_{x}$ is performed on Bell state $\alpha 
$ followed by a BXOR performed on both states, as Bell state $\alpha $ being
the source and Bell state $\beta $ being the target. Based on the
requirement for the transformation described in (14) and the unique
structure of the matrix ${\bf I}_{x}$, the series of row operations is
described stepwise in what follows. It then will be found that the present
method is a systematic one for solving the unknowns assumed in the matrix $%
{\bf M}_{w}$ (13).

In the first stage of row operations, we are confined to performing a
transformation of the matrix ${\bf M}_{w}$ (13) such that $m_{11}\rightarrow
e$ and $m_{1k},$ $m_{k1}\rightarrow 0$, $k=2,$ $3,$ $4,$and $5$, according
to the structure of $I_{x}.$ Based on an extension of the requirement for
the row operations (14), let $\det (m_{11})=1$ and $\det (m_{21})=...=\det
(m_{51})=0,$ which imply

\begin{equation}
a_{1}b_{2}\oplus a_{2}b_{1}=1,\text{ }c_{2}=0,\text{ }e_{1}=0,\text{ and }%
c_{1},\text{ }d_{1},\text{ }d_{2},\text{ }e_{2},f_{1},\text{ }f_{2}\in \{0,%
\text{ }1\}.
\end{equation}%
Clearly, there are totally 384 solutions for the unknowns appearing in (13)
to be considered in this stage. (6 for the condition $a_{1}b_{2}\oplus
a_{2}b_{1}=1$, 2 for each of the 6 arbitrary Boolean valued unknowns, and
thus totally $6\times 2^{6}=384$ solutions) To show the systematic way of
creating Boolean functions, however, only one among these 384 cases is
considered. To proceed, let us consider the case in which

\begin{equation}
a_{1}=b_{2}=1,\text{ }%
a_{2}=b_{1}=c_{1}=c_{2}=d_{1}=d_{2}=e_{1}=e_{2}=f_{1}=f_{2}=0,
\end{equation}%
Then simply by performing the operations shown in Fig. 2(a), we have the
transformation ${\bf M}_{w}\rightarrow {\bf M}_{w}^{\prime },$

\begin{eqnarray}
{\bf M}_{w}^{\prime } &=&\left[ 
\begin{array}{cccccccccc}
1 & 0 & 0 & 0 & 0 & 0 & 0 & 0 & 0 & 0 \\ 
0 & 1 & 0 & 0 & 0 & 0 & 0 & 0 & 0 & 0 \\ 
0 & 0 & c_{3} & c_{4} & c_{5} & c_{6} & c_{7} & c_{8} & c_{9} & c_{10} \\ 
0 & 0 & a_{3} & 1\oplus a_{4} & a_{5} & a_{6} & a_{7} & 1\oplus a_{8} & 
1\oplus a_{9} & a_{10} \\ 
0 & 0 & d_{3} & d_{4} & d_{5} & d_{6} & d_{7} & d_{8} & d_{9} & d_{10} \\ 
0 & 0 & 1 & 0 & 1 & 1 & 1 & 1 & 1 & 0 \\ 
0 & 0 & e_{3} & e_{4} & e_{5} & e_{6} & e_{7} & e_{8} & e_{9} & e_{10} \\ 
0 & 0 & 0 & 0 & 0 & 1 & 1 & 0 & 1 & 0 \\ 
0 & 0 & f_{3} & f_{4} & f_{5} & f_{6} & f_{7} & f_{8} & f_{9} & f_{10} \\ 
0 & 0 & 0 & 1 & 1 & 0 & 1 & 0 & 0 & 1%
\end{array}%
\right]  \nonumber \\
&=&\left[ 
\begin{array}{ccccc}
{\bf e} & 0 & 0 & 0 & 0 \\ 
0 & m_{22}^{\prime } & m_{23}^{\prime } & m_{24}^{\prime } & m_{25}^{\prime }
\\ 
0 & m_{32}^{\prime } & m_{33}^{\prime } & m_{34}^{\prime } & m_{35}^{\prime }
\\ 
0 & m_{42}^{\prime } & m_{43}^{\prime } & m_{44}^{\prime } & m_{45}^{\prime }
\\ 
0 & m_{52}^{\prime } & m_{53}^{\prime } & m_{54}^{\prime } & m_{55}^{\prime }%
\end{array}%
\right]
\end{eqnarray}%
and the equations

\begin{eqnarray}
a_{r}\oplus e_{r} &=&0, \\
b_{r}\oplus c_{r} &=&0,\text{ }r=3,\text{ }4,\text{ }...,\text{ }10.
\end{eqnarray}%
Here eqs. (18) and (19) are derived from the zero elements in the first row
of the $5\times 5$ ${\bf M}_{w}^{\prime }$ (17). So, we now have, in columns
2 to 5 of the $5\times 5$ ${\bf M}_{w}^{\prime }$ (17), the unknowns $c_{r}$%
, $d_{r}$,.$e_{r}$, and, $f_{r}$ remained to be solved because $a_{r}=e_{r}$
and $b_{r}=c_{r}$, from eqs. (18) and (19). At the end of this stage, we
should isolate Bell state 1 from being influenced by the following actions,
i.e., we from now on should maintain the first row and column of the $%
5\times 5$ ${\bf M}_{w}^{\prime }.$ We emphasize here that the other 383
solutions can be analyzed following the same way of obtaining the resulting
matrix ${\bf M}_{w}^{\prime }$ (17) in this stage. Of course, we can also
obtain multi-solution cases rather than \ the present cases if we
interchange the first column and the others in the assumed $5\times 5$
matrix ${\bf M}_{w}$ (13) and follow the same procedure of the first stage.
But this is not an important issue to be considered in this work, for we are
presenting a systematic method of creating Boolean functions involved in the
present QECC.

In the second stage of row operations, we will consider the general cases
that one of the elements in the second column of the $5\times 5$ ${\bf M}%
_{w}^{\prime }.$ (17) has determinant of unity and the other elements in the
same column have zero determinants. Three cases therefore need to be
considered in this stage. These cases are

\begin{eqnarray}
\text{(A) \ \ }c_{3}(1\oplus e_{4})\oplus c_{4}e_{3} &=&1,\text{ }d_{4}=0,%
\text{ }f_{3}=0,\text{ and }d_{3},\text{ }d_{4}\in \{0,\text{ }1\}; \\
\text{(B) \ \ }c_{3}(1\oplus e_{4})\oplus c_{4}e_{3} &=&0,\text{ }d_{4}=1,%
\text{ }f_{3}=0,\text{ and }d_{3},\text{ }d_{4}\in \{0,\text{ }1\}; \\
\text{(C) \ \ }c_{3}(1\oplus e_{4})\oplus c_{4}e_{3} &=&0,\text{ }d_{4}=0,%
\text{ }f_{3}=1,\text{ and }d_{3},\text{ }d_{4}\in \{0,\text{ }1\}.
\end{eqnarray}%
In these cases, there are 104 possible solutions for the unknowns appearing
in the above equations. We should, however, remember that these solutions
only belong to the one case considered in the first stage. We also note here
that if we interchange in the preceding stage the operating order of the two
BXOR's \ and their accompanying $B_{y}$'s shown in Fig. 2(a), then in the
present stage we will instead have only 72 possible solutions grouped into 4
cases. We therefore should consider the most general cases shown in eqs
(20)-(22) in determining the assumed matrix ${\bf M}_{w}$. Let us now
consider the solution, denoted by (A1), in which

\begin{equation}
\text{(A1) \ \ }c_{3}=1\text{ and }%
c_{4}=d_{3}=d_{4}=e_{3}=e_{4}=f_{3}=f_{4}=0,
\end{equation}%
to continue our analysis. Under this solution, we then perform the
operations shown in Fig. 2(b), obtaining the resulting matrix

\begin{eqnarray}
{\bf M}_{w}^{\prime \prime } &=&\left[ 
\begin{array}{cccccccccc}
1 & 0 & 0 & 0 & 0 & 0 & 0 & 0 & 0 & 0 \\ 
0 & 1 & 0 & 0 & 0 & 0 & 0 & 0 & 0 & 0 \\ 
0 & 0 & 1 & 0 & 0 & 0 & 0 & 0 & 0 & 0 \\ 
0 & 0 & 0 & 1 & 0 & 0 & 0 & 0 & 0 & 0 \\ 
0 & 0 & 0 & 0 & d_{5} & d_{6} & d_{7} & d_{8} & d_{9} & d_{10} \\ 
0 & 0 & 0 & 0 & 1 & 1 & 1 & 1 & 1 & 0 \\ 
0 & 0 & 0 & 0 & e_{5} & e_{6} & e_{7} & e_{8} & e_{9} & e_{10} \\ 
0 & 0 & 0 & 0 & 0 & 1 & 1 & 0 & 1 & 0 \\ 
0 & 0 & 0 & 0 & f_{5} & f_{6} & f_{7} & f_{8} & f_{9} & f_{10} \\ 
0 & 0 & 0 & 0 & 1\oplus e_{5} & e_{6} & 1\oplus e_{7} & 1\oplus e_{8} & 
1\oplus e_{9} & 1\oplus e_{10}%
\end{array}%
\right]  \nonumber \\
&=&\left[ 
\begin{array}{ccccc}
{\bf e} & 0 & 0 & 0 & 0 \\ 
0 & {\bf e} & 0 & 0 & 0 \\ 
0 & 0 & m_{33}^{\prime \prime } & m_{34}^{\prime \prime } & m_{35}^{\prime
\prime } \\ 
0 & 0 & m_{43}^{\prime \prime } & m_{44}^{\prime \prime } & m_{45}^{\prime
\prime } \\ 
0 & 0 & m_{53}^{\prime \prime } & m_{54}^{\prime \prime } & m_{55}^{\prime
\prime }%
\end{array}%
\right]
\end{eqnarray}

and the equations%
\begin{eqnarray}
c_{r}\oplus f_{r} &=&0,\text{ }r=5,\text{ }6,\text{ }...,\text{ }10 \\
d_{5} &=&e_{5},\text{ }d_{6}=e_{6},\text{ }d_{7}=e_{7},\text{ }d_{8}\oplus
e_{8}=1,\text{ }d_{9}\oplus e_{9}=1,\text{ }d_{10}=e_{10}.
\end{eqnarray}

Here, again, eqs. (25) and (26) are derived from the zero elements in the
second row of the $5\times 5$ matrix $M_{w}^{\prime \prime }$ (24). Using
eqs. (25) and (26), we now have only the remaining unknowns $e_{r}$ and $%
f_{r}$, $r=5,$ $6,$ $...,$ $10,$ to solve , so only one additional stage of
row operations is needed in what follows. At the end of this stage, we
should isolate both Bell state 1 and 2 from being influenced by the
following actions.

Following the same procedure as in the above stages, in the final stage, we
can perform transformations grouped into three cases and solve the remaining
unknowns subject to the corresponding final operations, respectively. These
three cases of transformation include

\begin{eqnarray}
\text{(A1}\alpha \text{) \ \ }m_{33}^{\prime \prime } &\rightarrow &e\text{
and }m_{43}^{\prime \prime },\text{ }m_{53}^{\prime \prime },\text{ }%
m_{34}^{\prime \prime },\text{ }m_{35}^{\prime \prime }\rightarrow 0\text{ };
\\
\text{(A1}\beta \text{) \ \ }m_{43}^{\prime \prime } &\rightarrow &e\text{
and }m_{33}^{\prime \prime },\text{ }m_{53}^{\prime \prime },\text{ }%
m_{44}^{\prime \prime },\text{ }m_{45}^{\prime \prime }\rightarrow 0\text{ };
\\
\text{(A1}\gamma \text{) \ \ }m_{53}^{\prime \prime } &\rightarrow &e\text{
and }m_{33}^{\prime \prime },\text{ }m_{43}^{\prime \prime },\text{ }%
m_{54}^{\prime \prime },\text{ }m_{55}^{\prime \prime }\rightarrow 0.
\end{eqnarray}%
Then, when eqs. (25) and (26) are used, we should perform the
transformations in cases (A1$\alpha $), (A1$\beta $), and (A1$\gamma $)
under the constraints 
\begin{equation}
e_{5}\oplus e_{6}=1,\text{ }e_{5}=0,\text{ }f_{5}e_{6}\oplus f_{6}(1\oplus
e_{5})=0,\text{ for case (A1}\alpha \text{);}
\end{equation}

\begin{equation}
e_{5}\oplus e_{6}=0,\text{ }e_{5}=1,\text{ }f_{5}e_{6}\oplus f_{6}(1\oplus
e_{5})=0,\text{ for case (A1}\beta \text{);}
\end{equation}

\begin{equation}
e_{5}\oplus e_{6}=0,\text{ }e_{5}=0,\text{ }f_{5}e_{6}\oplus f_{6}(1\oplus
e_{5})=1,\text{ for case (A1}\gamma \text{),}
\end{equation}%
respectively. Each of the above equations leads to two final results. We
shall denote them by

\begin{equation}
\text{(A1}\alpha 1\text{) \ \ }e_{5}=0,\text{ }e_{6}=1,\text{ }f_{5}=f_{6}=0;
\end{equation}

\begin{equation}
\text{(A1}\alpha 2\text{) \ \ }e_{5}=0,\text{ }e_{6}=1,\text{ }f_{5}=f_{6}=1;
\end{equation}

\begin{equation}
\text{(A1}\beta 1\text{) \ \ }e_{5}=e_{6}=1,\text{ }f_{5}=f_{6}=0;
\end{equation}

\begin{equation}
\text{(A1}\beta 2\text{) \ \ }e_{5}=e_{6}=1,\text{ }f_{5}=0,\text{ }f_{6}=1;
\end{equation}

\begin{equation}
\text{(A1}\gamma 1\text{) \ \ }e_{5}=e_{6}=0,\text{ }f_{5}=0,\text{ }f_{6}=1;
\end{equation}

\begin{equation}
\text{(A1}\gamma 2\text{) \ \ }e_{5}=e_{6}=0,\text{ }f_{5}=f_{6}=1,
\end{equation}%
respectively. Again, if we in the second stage of row operations interchange
the operating order of the two BXOR's shown in Fig.2(b), then we will
instead in this stage have 5 solutions subgrouped into 3 cases. The solution
of case (A1$\beta 2$) will be missed in this situation, so we should
consider the general cases given by (30)-(32), which lead to the 6 solutions
(33)-(38). Now, let us consider the case (A1$\alpha 1$). Using the result
shown in (33), we first perform the transformations $m_{33}^{\prime \prime
}\rightarrow e$ and $m_{43}^{\prime \prime },$ $m_{53}^{\prime \prime
}\rightarrow 0$ subject to suitable corresponding operations, then when
letting the transformations $m_{34}^{\prime \prime },$ $m_{35}^{\prime
\prime }\rightarrow 0$ to occur we obtain the final result for the remaining
unknowns $e_{r}$ and $f_{r}$ for $r=7,$ $8,$ $9,$ 10. The final result
itself then leads to the unique final transformations $m_{44}^{\prime \prime
},$ $m_{55}^{\prime \prime }\rightarrow e$ and $m_{54}^{\prime \prime },$ $%
m_{45}^{\prime \prime }\rightarrow 0$ when suitable operations are performed
correspondingly. The suitable operations performed in this stage are shown
in Fig. 2(c) and the final result reads

\begin{eqnarray}
e_{7} &=&0,\text{ }e_{8}=1,\text{ }e_{9}=0,\text{ }e_{10}=0,  \nonumber \\
\text{ }f_{1} &=&0,\text{ }f_{8}=0,\text{ }f_{9}=1,\text{ }f_{10}=0,
\end{eqnarray}%
for case (A1$\alpha 1$).

Summarizing the results (39), (33), (23), (16), incorporated with (25),
(26), (18), and (19), we now can construct the Boolean function for the case
(A1$\alpha 1$) at the end of the final stage of row operations, in which the
obtained matrix $M_{w}$ is expressed by

\begin{equation}
{\bf M}_{w(A1\alpha 1)}=\left[ 
\begin{array}{cccccccccc}
1 & 0 & 0 & 0 & 0 & 1 & 0 & 1 & 0 & 0 \\ 
0 & 1 & 1 & 0 & 0 & 0 & 0 & 0 & 1 & 0 \\ 
0 & 0 & 1 & 0 & 0 & 0 & 0 & 0 & 1 & 0 \\ 
1 & 0 & 0 & 1 & 0 & 0 & 0 & 1 & 1 & 0 \\ 
0 & 0 & 0 & 0 & 0 & 1 & 0 & 0 & 1 & 0 \\ 
0 & 0 & 1 & 0 & 1 & 1 & 1 & 1 & 1 & 0 \\ 
0 & 0 & 0 & 0 & 0 & 1 & 0 & 1 & 0 & 0 \\ 
0 & 1 & 0 & 0 & 0 & 1 & 1 & 0 & 1 & 0 \\ 
0 & 0 & 0 & 0 & 0 & 0 & 0 & 0 & 1 & 0 \\ 
0 & 0 & 0 & 1 & 1 & 0 & 1 & 0 & 0 & 1%
\end{array}%
\right] ,
\end{equation}%
which is exactly the same as the one given by Bennett et al.\cite{6} (shown
in eq. (77) of their article). A whole sequence of basic operations, as
shown in Fig. 3(a), is obtained by combining the three sub-sequences shown
in Figs. 2(a)-(c), which will transform the matrix ${\bf M}_{w(A1\alpha 1)}$
(40) into the matrix ${\bf I}_{x}$ expressed by

\begin{equation}
{\bf I}_{x}=\left[ 
\begin{array}{ccccc}
\begin{array}{cc}
1 & 0 \\ 
0 & 1%
\end{array}
&  &  &  &  \\ 
& 
\begin{array}{cc}
1 & 0 \\ 
0 & 1%
\end{array}
&  & 0 &  \\ 
&  & 
\begin{array}{cc}
0 & 1 \\ 
1 & 0%
\end{array}
&  &  \\ 
& 0 &  & 
\begin{array}{cc}
1 & 1 \\ 
0 & 1%
\end{array}
&  \\ 
&  &  &  & 
\begin{array}{cc}
1 & 0 \\ 
0 & 1%
\end{array}%
\end{array}%
\right] .
\end{equation}%
Reading the sequence of operations shown in Fig. 3(a), and all in other
figures, we should keep in mind that a ''backward''transformation ${\bf M}%
_{w}\rightarrow {\bf I}_{x}$ \ is accomplished by performing the operations
in the order from left to right, while a '' forward '' transformation ${\bf I%
}_{x}\rightarrow {\bf M}_{w}$ is implied to be undertaken by performing the
operations exactly in the reverse order, i.e., from right to left.

The rule of row operations on a given matrix ${\bf M}_{w}$, such as the one
shown in (40), however, indicates that we can transform the given matrix $%
{\bf M}_{w}$ in several ways into its corresponding ${\bf I}_{x}$'s,
implying there are several sequences of operations which can convert the
same given matrix ${\bf M}_{w}$ to distinct ${\bf I}_{x}$'s. Most
importantly, this also implies that we can seek a best sequence of
operations which consists of either the least number of total operations or
the least number of the BXOR operations alone, under a fixed correspondence
between the unmeasured states (the codewords in the first two rows of the
given $10\times 10$ ${\bf M}_{w}$ ) and the prescribed measurement result
vectors (embedded in ${\bf M}_{w}$ ) obtained from the given ${\bf M}_{w}$.
So, the final step of our method is to seek a shortened sequence of
operations according to the rule of row operations on the matrix ${\bf M}%
_{w} $ obtained in the preceding step. This step begins with the
''starting''sequence of operations constructed simply by combining the three
sub-sequences used in the preceding step in obtaining a specific matrix $%
{\bf M}_{w}$ for a successful error-correction. A systematic way to shorten
the starting sequence of operations, is to re-display the complete row
operations in transforming the specific ${\bf M}_{w}$ into its corresponding 
${\bf I}_{x}$'s under all the permutations of the BXOR's appearing in the
starting sequence. We show in Fig. 3(b) an example of the shortened sequence
of operations involved in transforming the matrix ${\bf M}_{w}$ shown in
(40) into the matrix

\begin{equation}
{\bf I}_{x}=\left[ 
\begin{array}{ccccc}
\begin{array}{cc}
1 & 0 \\ 
0 & 1%
\end{array}
&  &  &  &  \\ 
& 
\begin{array}{cc}
0 & 1 \\ 
1 & 0%
\end{array}
&  & {\LARGE 0} &  \\ 
&  & 
\begin{array}{cc}
0 & 1 \\ 
1 & 0%
\end{array}
&  &  \\ 
& {\LARGE 0} &  & 
\begin{array}{cc}
0 & 1 \\ 
1 & 1%
\end{array}
&  \\ 
&  &  &  & 
\begin{array}{cc}
1 & 0 \\ 
1 & 1%
\end{array}%
\end{array}%
\right]
\end{equation}%
Clearly, the new version shown in Fig 3 (b) is a result obtained when we
interchange the first two BXOR's in the original sequence of operations
shown in Fig. 3(a); It consists of only ten operations in which seven BXOR's
are required. The two sequences shown in Fig. 3 are of course different,
resulting in distinct ${\bf I}_{x}$'s, as shown in (41) and (42), but they
are equivalent because they both lead to the same correspondence between the
unmeasured states and the prescribed measurement result vectors, as
tabulated in Table 2 and named by case (A1$\alpha 1$).

\section{Results and Discussion}

In the preceding section we have shown the complete procedure for creating
Boolean functions involved in the present 5-EPR pair, single
-error-correcting code. We have given an example of the Boolean functions
derived by using the present method, namely, the given matrix ${\bf M}_{w}$
in (40) and its best corresponding sequence of operations shown in Fig.
3(b). To show more examples, we have also analyzed the cases (A1$\alpha 2$),
(A1$\beta 1$), (A1$\beta 2$), (A1$\gamma 1$), and (A1$\gamma 2$), in which
the choices (16), (23), and (33)-(38) are taken, respectively. Their final
correspondences between the unmeasured Bell states and the measurement
results are shown in Table 2 and the gate arrays of the corresponding
sequences of operations are shown in Fig. 4. It is found that the sequence
of operations shown in Fig. 4(e) for the case (A1$\gamma 2$) contains only
six BXOR's, while the others, including the one shown in Fig. 3(b) and those
shown in Figs. 4(a)-(d), all contain seven BXOR's. To perform a more
reliable error-correcting protocol, we may require to construct a Boolean
function under the effect of a sequence \ of operations containing fewer
BXOR's since two-bit operations could be more difficult ones to implement in
a physical apparatus \cite{19}, so the sequence of operations for the case
(A1$\gamma 2$) may be crucially important. Indeed, there are many Boolean
functions in which the sequences of operations contain only six BXOR's.
Using our systematic method, it becomes relatively easier than using the
numerical Monte Carlo method to find out such Boolean functions. In
principle, it can be easily done to find sequences containing only six BXOR
operations if we can skillfully choose suitable solutions and apply
operations in suitable order so that more zero elements are present both in
the second column of the $5\times 5$ matrix ${\bf M}_{w}^{\prime }$ (17) and
the third column of $5\times 5$ ${\bf M}_{w}^{^{\prime }\prime }$\ matrix
(24). For instance, if we maintain the same choice (16) in the first stage
of row operations, then choose the solution in which $%
c_{3}=c_{4}=d_{3}=d_{4}=e_{3}=e_{4}=f_{4}=0$ and $f_{3}=1$ as one of the
cases (C) in the second stage of row operations, and finally choose $%
e_{5}=e_{6}=f_{5}=f_{6}=0$ in the final stage of row operations, we
eventually will construct a Boolean function in which the matrix ${\bf M}%
_{w} $ is given by

\begin{equation}
{\bf M}_{w(\text{C}1\beta 1)}=\left[ 
\begin{array}{cccccccccc}
1 & 0 & 0 & 0 & 0 & 0 & 1 & 0 & 0 & 1 \\ 
0 & 1 & 0 & 0 & 0 & 0 & 0 & 1 & 1 & 1 \\ 
0 & 0 & 0 & 0 & 0 & 0 & 0 & 1 & 1 & 1 \\ 
1 & 0 & 0 & 1 & 0 & 0 & 0 & 1 & 1 & 0 \\ 
0 & 0 & 0 & 0 & 1 & 0 & 1 & 0 & 0 & 1 \\ 
0 & 0 & 1 & 0 & 1 & 1 & 1 & 1 & 1 & 0 \\ 
0 & 0 & 0 & 0 & 0 & 0 & 1 & 0 & 0 & 1 \\ 
0 & 1 & 0 & 0 & 0 & 1 & 1 & 0 & 1 & 0 \\ 
0 & 0 & 1 & 0 & 0 & 0 & 0 & 1 & 1 & 1 \\ 
0 & 0 & 0 & 1 & 1 & 0 & 1 & 0 & 0 & 1%
\end{array}%
\right] ,
\end{equation}%
and the corresponding sequence of operations is shown in Fig. 5. It is found
that such a sequence of operations contains only six BXOR's and three other
operations; it is even better than the one in the case (A1$\gamma 2$) since
the latter contains six BXOR's but four other operations. The correspondence
between the measured Bell states and the measurement results of such Boolean
function is also shown in Table 2, denoted as case (C1$\beta 1$).

We have in this work presented some analytical results for the Boolean
function under the fixed designation of measurement results as given by (11)
and embedded in the assumed matrix shown by (13). If we name all of the
solutions that can be obtained under the present designation of measurement
results a solution group, then, following the same scenario in deducing the
Boolean function, we can possibly obtain other solution groups under
distinct arrangements of the measurement results. For example, based on the
representation of $5\times 5$ matrix for the assumed ${\bf M}_{w}$ shown in
(13), we in fact can choose any one of the five columns and locate it at the
first column in accord with the confinement in the first stage of row
operations interpreted in the preceding section. So, we conclude that there
are totally 5 ''independent'' solution groups under such arrangements that
can be obtained by using our method. Besides, possible designations of the
array of measurement results can also be obtained by performing row
operations on the given array (11). The solutions resulted from such
designations, however, are distinct but ''dependent'' to those resulted from
the original array (11) because they are distinct to each other only due to
effects of some BXOR operations. For example, as we can change the first row
of the array (11) by adding the second row to it and have a reduced array
given by

\begin{equation}
{\bf M}_{v}=\left[ 
\begin{array}{cccccccccc}
1 & 0 & 1 & 1 & 1 & 1 & 1 & 0 & 0 & 0 \\ 
0 & 0 & 1 & 0 & 1 & 1 & 1 & 1 & 1 & 0 \\ 
0 & 1 & 0 & 0 & 0 & 1 & 1 & 0 & 1 & 0 \\ 
0 & 0 & 0 & 1 & 1 & 0 & 1 & 0 & 0 & 1%
\end{array}%
\right] ,
\end{equation}%
we then have dependent matrices ${\bf M}_{w}$\ different from those resulted
from the designation (11) by only a BXOR operation acting on Bell states 2 (
as the target ) and 3 ( as the source ).

As a final result, there are so many possible Boolean functions subject to
the prescribed array of measurement results (11) that can be implemented in
the 5-EPR-pair, single-error-correcting code and they can be analytically
derived by our systematic method. If all these Boolean functions are to be
established, however, a computer program should be developed following the
present scenario to make it practical and feasible. The computer program to
be developed should be capable to determine the suitable matrices M$_{w},$\
then to reconstruct the best sequences of operations for the given matrices,
and finally help to select the best Boolean functions which can be performed
by least basic operations or least BXOR's. To develop the computer program,
however, is not an issue to be concerned with in this work.

\section{Conclusion}

In this work a systematic method has been presented for creating Boolean
functions required in the 5-EPR-pair, single-error-correcting code first
introduced by Bennett et al.\cite{6}. Although so far the 5-EPR-pair,
single-error-correcting code has not been undertaken experimentally, we here
have complemented the mathematical theory of the QECC by showing an
analytical way for creating the required Boolean function. Distinct to the
previously used Monte Carlo computer search, which may consume a lot of
trial and error efforts, the present method can help creating the Boolean
function in an analytical procedure. In the spirit of row operations on a
matrix used in the present method, we are also capable to establish the best
sequence of basic operations for every given Boolean function. We have given
some analytical results of the Boolean function and the associated sequence
of operations using the present method. Some of the results are better than
the others because they require fewer number of total operations or fewer
number of BXOR operations and then can make the QECC more reliable. The
three systematic steps for deducing the present results that have been
described in detail in the preceding sections have shown the scenario
helpful in creating the Boolean function that are potentially useful in the
5-EPR-pair QECC. The effort to create all the Boolean functions, however,
can be accomplished more efficiently if aided by a computer program designed
in accord with the present scenario for creating the Boolean
function.\bigskip 
\[
\begin{tabular}[t]{|l|l|l|}
\hline
$i$ & $x^{(i)}$ & $v^{(i)}$ \\ \hline
0 & 0000000000 & 0000 \\ \hline
1 & 1000000000 & 1000 \\ \hline
2 & 0100000000 & 0010 \\ \hline
3 & 1100000000 & 1010 \\ \hline
4 & 0010000000 & 0100 \\ \hline
5 & 0001000000 & 1001 \\ \hline
6 & 0011000000 & 1101 \\ \hline
7 & 0000100000 & 0101 \\ \hline
8 & 0000010000 & 0110 \\ \hline
9 & 0000110000 & 0011 \\ \hline
10 & 0000001000 & 0111 \\ \hline
11 & 0000000100 & 1100 \\ \hline
12 & 0000001100 & 1011 \\ \hline
13 & 0000000010 & 1110 \\ \hline
14 & 0000000001 & 0001 \\ \hline
15 & 0000000011 & 1111 \\ \hline
\end{tabular}%
\bigskip 
\]%
Table 1. One possible correspondence between the error syndrome $x^{(i)}$
and the measurement-result vectors $v^{(i)}.$ Note that the relations $%
x^{(3k-2)}\oplus x^{(3k-1)}=x^{(3k)}$ and $v^{(3k-2)}\oplus
v^{(3k-1)}=v^{(3k)},$ $k=1,2,...,5,$ are satisfied. \bigskip

\[
\begin{tabular}[t]{|l|l|l|l|l|l|l|l|l|}
\hline
$i$ & $v^{(i)}$ & $w^{^{\prime }(i)}$(A1$\alpha 1$) & (A1$\alpha 2$) & (A1$%
\beta 1$) & (A1$\beta 2$) & (A1$\gamma 1$) & (A1$\gamma 2$) & (C1$\beta 1$)
\\ \hline
0 & 0000 & 00 & 00 & 00 & 00 & 00 & 00 & 00 \\ \hline
1 & 1000 & 10 & 10 & 10 & 10 & 10 & 10 & 10 \\ \hline
2 & 0010 & 01 & 01 & 01 & 01 & 01 & 01 & 01 \\ \hline
3 & 1010 & 11 & 11 & 11 & 11 & 11 & 11 & 11 \\ \hline
4 & 0100 & 01 & 01 & 01 & 01 & 01 & 01 & 00 \\ \hline
5 & 1001 & 00 & 00 & 00 & 00 & 00 & 00 & 00 \\ \hline
6 & 1101 & 01 & 01 & 01 & 01 & 01 & 01 & 00 \\ \hline
7 & 0101 & 00 & 01 & 10 & 10 & 00 & 01 & 00 \\ \hline
8 & 0110 & 10 & 11 & 10 & 11 & 01 & 01 & 00 \\ \hline
9 & 0011 & 10 & 10 & 00 & 01 & 01 & 00 & 00 \\ \hline
10 & 0111 & 00 & 01 & 00 & 01 & 11 & 11 & 10 \\ \hline
11 & 1100 & 10 & 10 & 00 & 01 & 01 & 00 & 01 \\ \hline
12 & 1011 & 10 & 11 & 00 & 00 & 10 & 11 & 11 \\ \hline
13 & 1110 & 01 & 00 & 11 & 11 & 01 & 00 & 01 \\ \hline
14 & 0001 & 00 & 01 & 00 & 01 & 11 & 11 & 11 \\ \hline
15 & 1111 & 01 & 01 & 11 & 10 & 10 & 11 & 10 \\ \hline
\end{tabular}%
\bigskip 
\]

Table 2. The correspondences between the measurement results, denoted by $%
v^{(i)},$ and the unmeasured Bell states, coded as $w^{^{\prime }(i)},$ for
the cases (A1$\alpha 1$) to (C1$\beta 1$). The notation $i$ enumerates the
error syndromes in the QECC.

\bigskip

\bigskip {\LARGE Figure caption.}

\bigskip

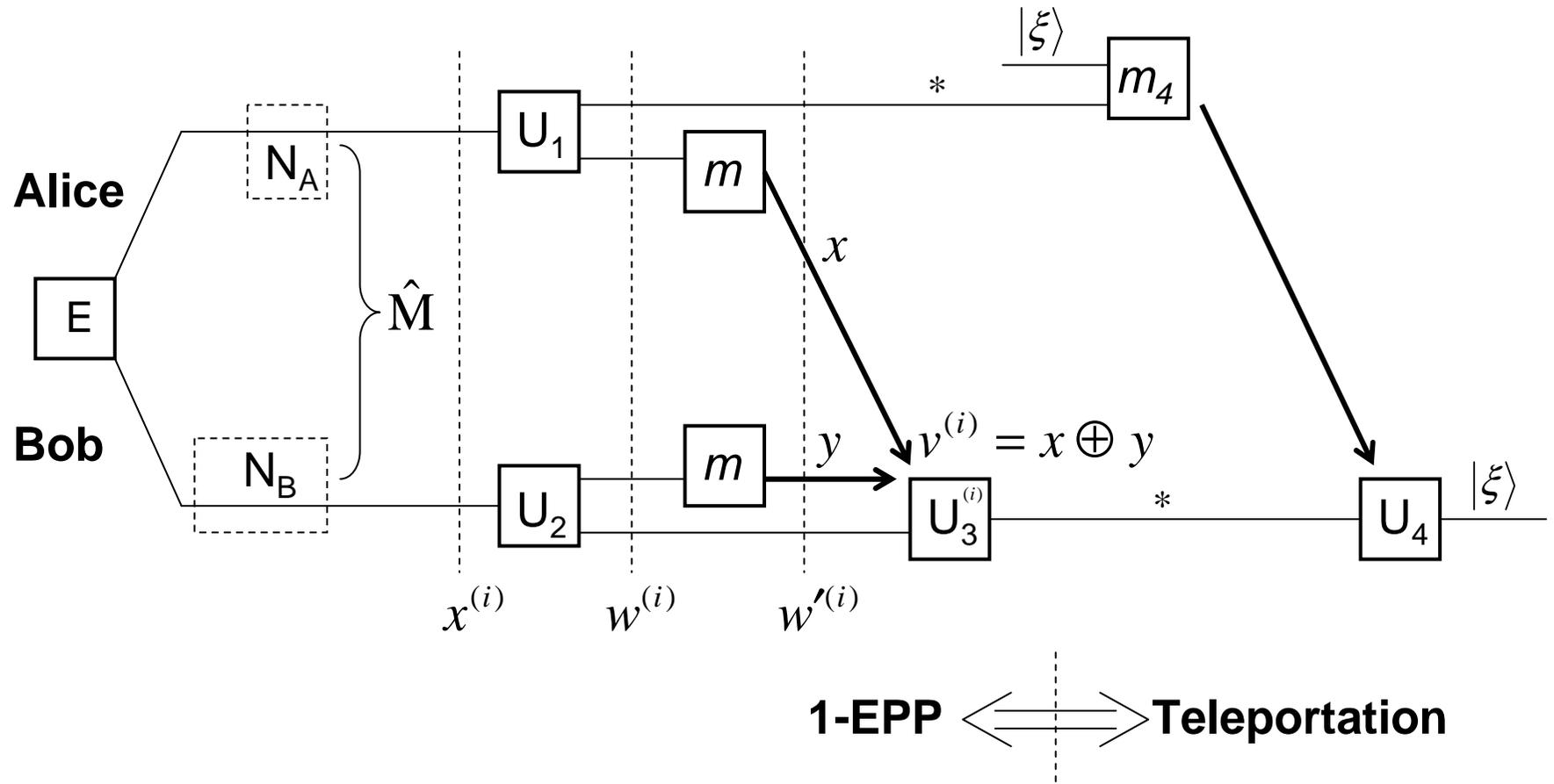
Fig. 1. The 5-EPR-pair, single-error-correcting code is derived from the
combination of the 1-EPR and teleportation schematically shown in this
figure, with the notations used in the context. The 1-EPR results in
perfectly entangled state (*) which are then used to teleport $\left| \xi
\right\rangle $ safely from Alice to Bob, completing the QECC. The
teleportation is initiated with Alice's Bell measurement $M_{4}$\ and is
completed by Bob's unitary transformation U$_{4}.$

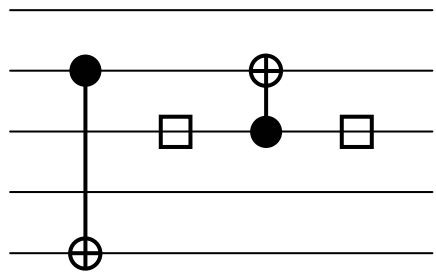
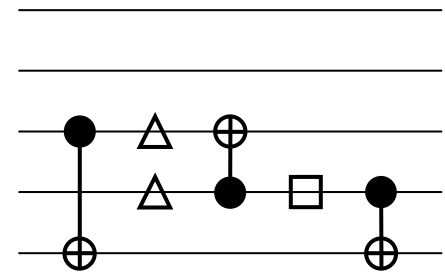
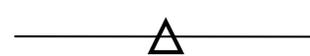
Fig. 2. The three quantum gate arrays performed in the stages of row
operations: (a) for ${\bf M}_{w}\rightarrow {\bf M}_{w}^{\prime }$ ; (b) for 
${\bf M}_{w}^{\prime }\rightarrow {\bf M}_{w}^{\prime \prime }$ ; and (c)
for ${\bf M}_{w}^{\prime \prime }\rightarrow {\bf I}_{x}$ .

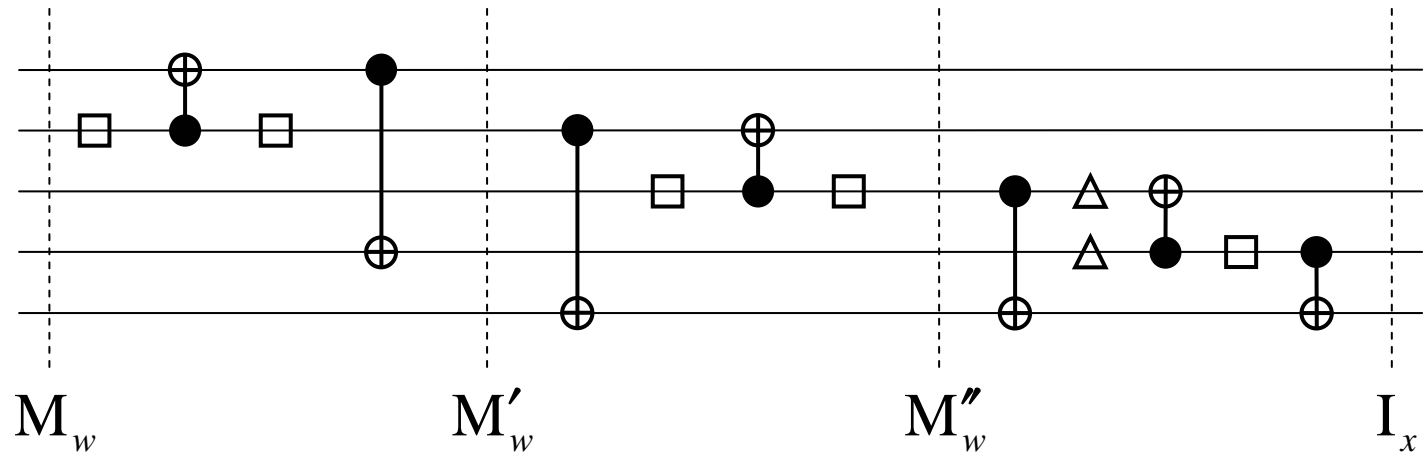
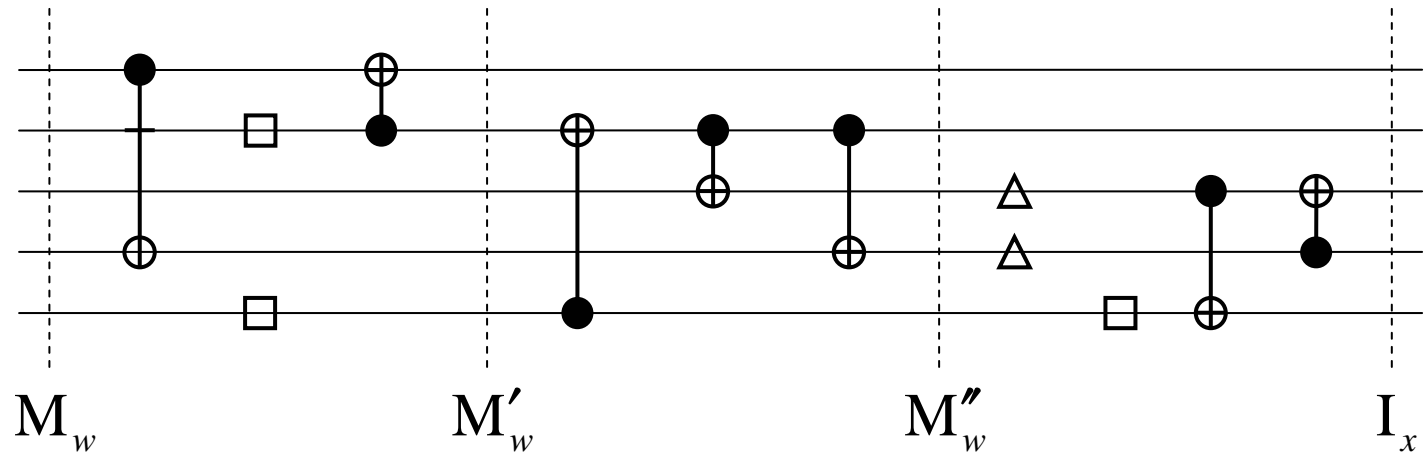
Fig. 3. The quantum gate arrays used in the case (A1$\alpha 1$) for
transforming the matrix ${\bf M}_{w(\text{A}1\alpha 1)}$ into (a) ${\bf I}%
_{x}$ (41) and (b) ${\bf I}_{x}$ (42), respectively. In realistic
situations, Alice and Bob should perform the sequences of operations in the
direction from right to left.

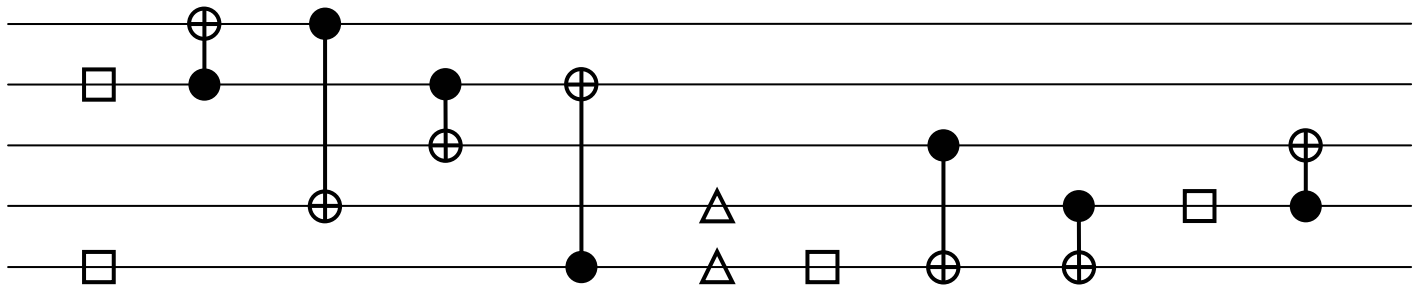
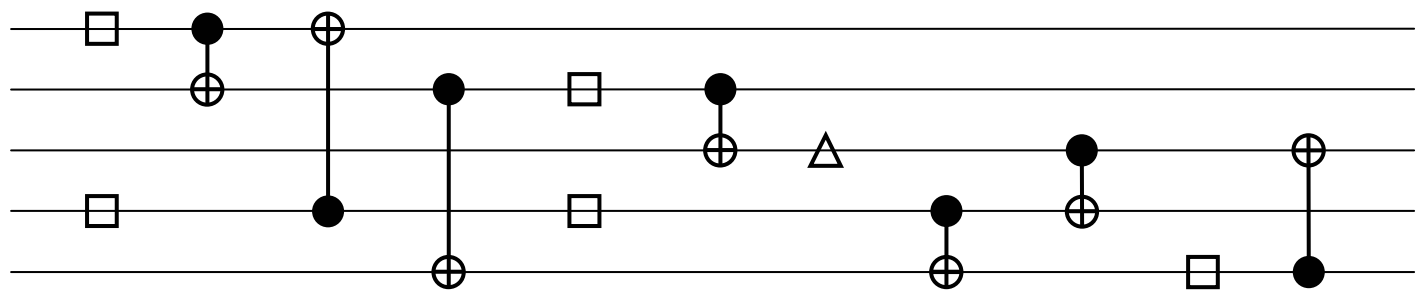
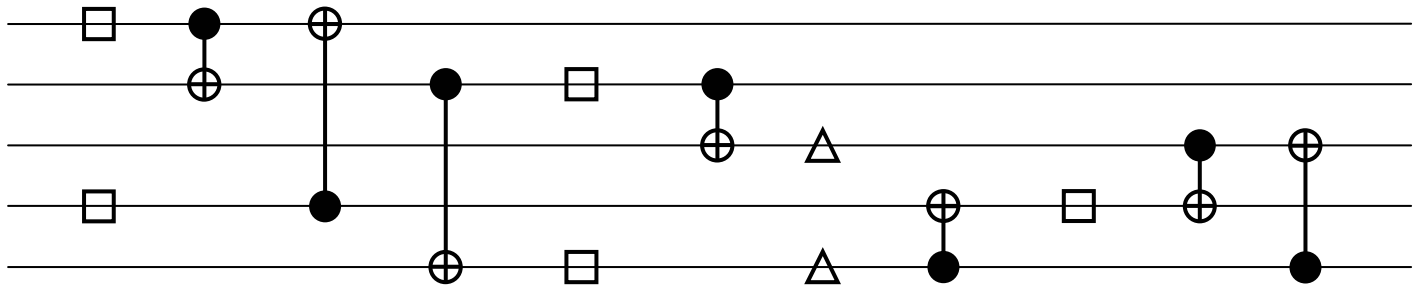
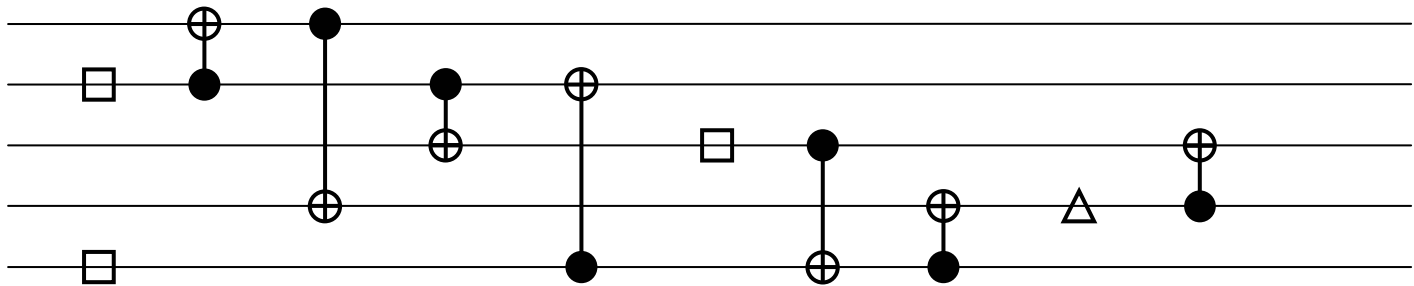
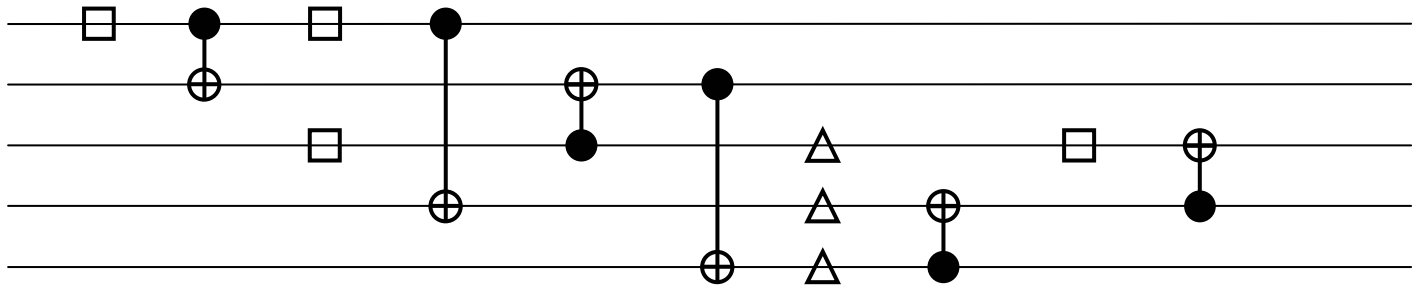
Fig 4. The quantum gate arrays which may be used in the cases (A1$\alpha 2$%
), (A1$\beta 1$), (A1$\beta 2$), (A1$\gamma 1$), and (A1$\gamma 2$).

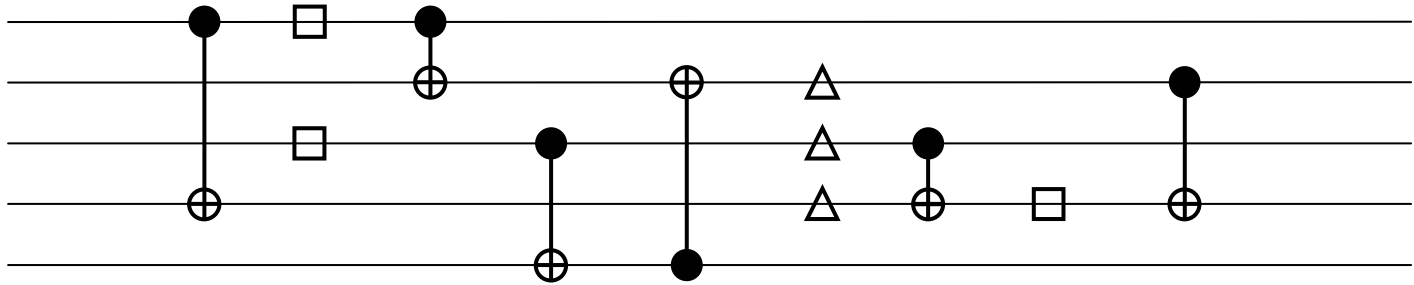
Fig. 5. The quantum gate array used in the case (C1$\beta 1$). This sequence
of operations contains only six BXOR's and three other operations.


\begin{thebibliography}{99}
\bibitem{1} C. H. Bennett, G. Brassard, C. Crepeau, R. Jozsa, A. Peres, and
W. K. Wootters, ''Teleporting an unknown quantum state via dual classical
and Einstein-Podolsky-Rosen channels,'' {\it Phys. Rev. Lett}., vol. 70, no.
13, pp. 1895-1899, 1993.

\bibitem{2} B. Schumacher, ''Quantum Coding, '' {\it Phys. Rev. A}, vol. 51,
no. 4, pp. 2738-2747, 1995.

\bibitem{3} C. H. Bennett and S. J. Wiesner, ''Communication via One- and
Two-particle Operators on Einstein-Podolsky-Rosen States,'' {\it Phys. Rev.
Lett, }vol{\it . 69}, pp. 2881-2884, 1992.

\bibitem{4} C. H. Bennett, G. Brassard, S. Popescu, B. Schumacher, J. A.
Smolin, and W. K. Wootters, ''Purification of Noisy Entanglement and
Faithful Teleportation via Noisy Channels,'' {\it Phys. Rev. Lett.}$,$ vol.
76, no.5, pp. 722-725 ,1996.

\bibitem{5} D. Deutsch, A. Ekert, R. Jozsa, C. Macchiavello, S. Popescu, and
A. Sanpera, ''Quantum Privacy Amplification and the Security of Quantum
Cryptography over Noisy Channels,'' {\it Phys. Rev. Lett}., vol. 77, no. 13,
pp. 2818-2821,1996.

\bibitem{6} C. H. Bennett, D. P. DiVincenzo, J. A. Smolin, and W. K.
Wootters, ''Mixed-state Entanglement and Quantum Error Correction, '' {\it %
Phys. Rev. A},{\it \ }vol. 54, no. 5, pp. 3824-3851, 1996.

\bibitem{7} P. W. Shor and J. A. Smolin, ''Quantum Error-Correcting Codes
Need Not Completely Reveal the Error Syndrome,''Report No. quant-ph/9604006.

\bibitem{8} A. R. Calderbank and P. W. Shor, ''Good Quantum Error-Correcting
Codes Exist, ''{\it Phys. Rev. A} vol. 54, pp. 1098-1105, 1996.

\bibitem{9} A. M. Steane, ''Error Correcting Codes in Quantum Theory, '' 
{\it Phys. Rev. Lett.}$,$ vol. 76, pp. 793-797, 1996.

\bibitem{10} Isaac L. Chuang and Yoshihisa Yamamoto, ''Quantum Bit
Regeneration,'' {\it Phys. Rev. Lett}, vol. 52, pp. 4281-4284, 1996.

\bibitem{11} P. W. Shor, ''Scheme for Reducing Decoherence in Quantum
Computer Memory, '' {\it Phys. Rev. A}, vol. 52, pp. 2493-2496, 1995.

\bibitem{12} Raymond Laflamme, Cesar Miquel, Juan Pablo Paz, and Wojciech
Hubert Zurek, ''Perfect Quantum Error Correcting Code, '' {\it Phys. Rev.
Lett}, vol. 77, pp. 198-201, 1996.

\bibitem{13} Artur Ekert and Chiara Macchiavello, ''Quantum Error Correction
for Communication, '' {\it Phys. Rev. Lett}, vol. 77, pp. 2585-2588, 1996.

\bibitem{14} Samuel L. Braunstein, ''Quantum Error Correction of Dephasing
in 3 Qubits, ''Report No. quant-ph/9603024.

\bibitem{15} M. B. Plenio, V. Vedral, and P. L. Knight, ''Conditional
Generation of Error Syndromes in Fault-tolerant Error Correction , '' {\it %
Phys. Rev. A}, vol. 55, pp. 4593-4596, 1997.

\bibitem{16} Lev Vaidman, Lior Goldenberg, and Stephen Wiesner, ''Error
Prevention Scheme with Four Particles, '' {\it Phys. Rev. A}, vol. 54, pp.
1745-1748, 1996.

\bibitem{17} Emanuel Knill and Raymond Laflamme, ''A Theory of Quantum
Error-Correcting Codes, '' Report No. quant-ph/9604034; Emanuel Knill1,
Raymond Laflamme and Lorenza Viola, ''Theory of Quantum Error Correction for
General Noise, '' {\it Phys. Rev. Lett}, vol. 84 pp. 2525-2528, 2000.

\bibitem{18} E. Knill, R. Laflamme, R. Martinez, and C. Negrevergne,
''Benchmarking Quantum Computers: The Five-Qubit Error Correcting Code, '' 
{\it Phys. Rev. Lett, }vol{\it . 86}, pp. 5811-5814, 2001.

\bibitem{19} Adriano Barenco, Charles H. Bennett, Richard Cleve, David P.
DiVincenzo, Norman Margolus, Peter Shor, Tycho Sleator, John A. Smolin, and
Harald Weinfurter, ''Elementary Gates for Quantum Computation, '' {\it Phys.
Rev. A}, vol. 52, pp. 3457-3467, 1995.
\end{thebibliography}
\end{document}